# Superconducting screening on different length scales in high-quality bulk MgB$_2$ superconductor

J. Horvat, S. Soltanian, A. V. Pan and X. L. Wang

*Institute for Superconducting and Electronic Materials, University of Wollongong, NSW 2522, Australia*

Field dependence of irreversible magnetic moment $\Delta m$ was obtained from magnetic hysteresis loops for a number of bulk MgB$_2$ superconductors. The field dependence of $\Delta m$ exhibits inflections at two characteristic fields, $H_t$ and $H_i$ ($H_t < H_i$). These two fields increase linearly with the logarithm of the sample size, each extrapolating to zero at a different characteristic sample size. Magneto-optical, optical and scanning electron microscopic examinations show that these extrapolated sample sizes match the sizes of the main microscopic features of the samples. The inflection at $H_t$ and the sample size dependence of $\Delta m$ and $H_t$ are associated with voids scattered through the samples, which are observed for all bulk MgB$_2$ samples. The voids encircle cells of MgB$_2$ material of tens of micrometers in size. The cells are connected by narrow bridges. The superconducting currents screening the whole of the sample have to cross these bridges and they are concentrated in them. This promotes additional superconducting screening around the cells, where the current density would otherwise be smaller than the critical current density $J_c$. The field dependence of the currents pertinent to each of the screenings follows the stretched exponential law. However, these two screening currents decrease with $H$ with a different rate, and $H_t$ is a field of the transition from the dominance of one of the currents in $\Delta m$ to the other. The screening that gives predominant contribution to $\Delta m$ for $H < H_t$ is confined to inside the cells, circulating on ~10 μm-scale. For $H_t < H < H_i$, the dominant screening currents circulate the whole of the sample, percolating through the bridges. If $J_c$ is calculated from the critical state model assuming only the screening around the whole of the sample, erroneous value of $J_c$ and its functional dependence on the field are obtained. This also leads to an artifact of the sample size-dependent $J_c$. However, because these two stretched exponential contributions decrease with $H$ with a different rate, they can be separated for most of the samples. The magnetically obtained $J_c$ for $H_t < H < H_i$ is defined by the screening currents percolating around the whole of the sample, and it is in good agreement with the transport $J_c$. There is also a third tier of superconducting screening on a micrometer scale, inside the cells. It is associated with clusters of superconducting crystals that make up the cells. However, its contribution to $\Delta m$ is negligible for $H < H_i$, which are the fields of technological interest.

I. INTRODUCTION

High quality $MgB_2$ pellets and wires exhibit large critical current densities, which are difficult to measure in a transport method. Because of this, most of the claims of achieving a high critical current density ($J_c$) and of improvement of vortex pinning are based on magnetic measurements. However, we will show that the value of magnetically obtained $J_c$ and its field dependence vary with the sample size and can be quite erroneous without proper analysis of the experimental data.

One of the main technological advantages of high-quality bulk $MgB_2$ is that its superconducting crystals are connected well. This results in the absence of weak links[1], which are the main weakness of high-temperature superconductors. For high-temperature superconductors, there is a sharp decrease of $J_c$ at low magnetic fields due to the breaking down of the weak links in the field and decoupling of the superconducting grains[2, 3]. The external magnetic field ($H$) penetrates between the grains where the weak links are decoupled.

Because of the absence of the weak links, this abrupt decrease of $J_c$ in low fields does not occur in the bulk $MgB_2$. Therefore, the field dependence of $J_c$ should be entirely defined by the vortex pinning and suppression of the superconducting order parameter as the field approaches $H_{c2}$. However, magnetic measurements of $J_c$ for bulk $MgB_2$ showed that the zero-field $J_c$, $J_{c0}$, strongly decreased with the sample volume[4] for samples smaller than about 10 mm$^3$. For larger samples, $J_{c0}$ did not change significantly with the sample volume anymore. These results indicated additional superconducting screening, on the length-scale smaller than the sample. The relative contribution of each of the screenings to the measured irreversible magnetic moment $\Delta m$ depended on the size of the sample when it approached the length-scale of the additional screening. Both of the screenings contributed to $\Delta m$, resulting in an artificial increase of magnetic $J_{c0}$ as the sample size approached the length-scale of the additional screening. The relative contribution of each of the screenings to $\Delta m$ did not change with the sample size for large samples.

This paper shows that the superconducting screening at different length scales has a strong influence on the field dependence of $\Delta m$. The rate of decrease of $\Delta m$ with field is shown to be different for each of the screenings. Characteristic fields are identified, representing the transition from the dominance of one of the screenings in d($\Delta m$)/d$H$ to the dominance of the other. The sample size dependence of these fields gives the value of the characteristic length-scales for different superconducting screenings. The field dependence of $\Delta m$ for each of the screening follows the stretched exponential law [5, 6], and the measured $\Delta m$ is the sum of these contributions. This implies that the field dependence of $J_c$, obtained from $\Delta m$ using the critical state model, is an artifact of different contributions to the measured $\Delta m$. The magnetically obtained $J_c$ can be much larger than the transport $J_c$ because of this artifact. Therefore, any conclusions derived from the magnetic measurements of $MgB_2$ should be treated with caution, despite that there is no weak links in high-quality $MgB_2$. This is of great concern, because the efforts to develop $MgB_2$ in the form of pellets and wires usually rely on magnetic measurements.

II. EXPERIMENTAL PROCEDURE

A number of different $MgB_2$ samples were measured, from our first $MgB_2$ pellets, [7] to the best $MgB_2$/Fe wires [8]. The pellets were prepared by mixing fine powders of magnesium and boron, with slight surplus of Mg. This mixture was pressed into pellets, which were sealed into iron tube and sintered in flowing high-purity argon. The sintering temperatures were between 700 and 850 °C and sintering time ranged from 1 minute to 1 hour. The iron-sheathed wires were prepared by first filling the powder into iron tubes, which were drawn to desired diameter, sealed into another iron tube, and then sintered in flowing high purity argon. The exact preparation procedure for these samples was described elsewhere [9, 10]. The iron sheath was removed before the magnetic measurements, to avoid its contribution to the signal and magnetic interaction between the iron and $MgB_2$ [11].

Another group of samples was prepared by hot isostatic pressing (HIP). The mixture of magnesium and boron powders was pressed at 150 MPa at 850 °C for 1 hour, in argon atmosphere. The density of the obtained $MgB_2$ pellet was 1.9 g/cm$^3$, as compared to ≈1.5 g/cm$^3$ for the pellets obtained at atmospheric pressure. The HIP-ed samples were used for accurate measurements of the effect of the sample size on the field dependence of $\Delta m$. They were cut into a rectangular shape, with magnetic field applied along the longest sample dimension. Each of the three sample dimensions were subsequently decreased by 20% after each of the measurements. In this way, the proportions of the sample remained the same for all the measurements, avoiding possible effects of the sample shape on the results. Exact dimensions of these samples can be found in Ref [4].

All the measured samples had critical temperature of about 38-39K, as obtained from the measurements of magnetic susceptibility. XRD analysis showed they consist of $MgB_2$, with less than 10 % of MgO.

The field dependence of $\Delta m$ was obtained from the magnetic hysteresis loops. Most of them were measured by a Quantum Design PPMS. Some of the loops were measured by an Oxford Instruments VSM, with vibration amplitude 1.5 mm and frequency 45 Hz. The results obtained from both of the instruments were consistent with each other. The sweep rate of magnetic field was 50 Oe/s.

The transport $J_c$ was measured for iron sheathed wires by pulsed current method. This method was necessary to avoid heating of the sample, because the critical currents were of several hundreds of amperes. The duration of the current-pulse was about 1 ms. The detailed description of the experimental set-up can be found elsewhere [11].

Magneto-optical imaging was performed by employing an epitaxial ferrite-garnet indicator film, with in-plane magnetization [12]. The sample was placed in a cryostat with optical window and magnetic pattern produced on the indicator film by magnetic fields above the sample surface was observed by a microscope. Magnetic field was produced by a solenoid with an iron core.

III. EXPERIMENTAL RESULTS

The field dependence of $J_c$ for $MgB_2$ is usually obtained from the measurements of magnetic hysteresis loops [7], using the critical state model [13, 14]. $J_c$ is in this model proportional to the measured irreversible magnetic moment $\Delta m$, which is the height of the magnetic hysteresis loop, and inversely proportional to the sample size. A typical field dependence of $\Delta m$ per unit volume of an $MgB_2$ sample is shown in Fig. 1, with log-lin scales. The same qualitative behavior was observed for HIP-ed samples, $MgB_2$ cores of Fe-sheathed wires, $MgB_2$ pellets, as well as for the $MgB_2$ powders with the particle size ranging from a few μm to tens of μm. Three different regimes in the field dependence of $\Delta m$ can be identified in Fig.1, with transitions between them at fields about $H = 1$ T and 4 T. The step at 4 T is the more obvious transition in Fig.1, because the logarithmic scale was used for $J_c$. However, the inflection at 1T occurs at $\approx$ 40% of the value of $\Delta m$ for H = 0. Therefore, the regime for $H <$ 1T accounts for $\approx$ 60% of the decrease of $\Delta m$ with $H$.

These three regimes can be much easier observed in the plots of the field dependence of the relative decrease of $\Delta m$ with $H$, -d($\Delta m$)/d$H$ * ($\Delta m$)$^{-1}$, which can be re-written as -d(ln($\Delta m$))/d$H$. This is shown in Fig. 2 for the same sample as in Fig.1. The transition fields at 1 and 4 T are identified as the peaks in Fig.2 and they are denoted as $H_t$ and $H_i$, respectively. For all measured samples, the data of -d(ln($\Delta m$))/d$H$ vs. $H$ are linearized in log-log plots for $H < H_t$ (Fig.2). The gradient of log[-d(ln($\Delta m$))/d$H$] vs. log[$H$] is denoted as $n-1$. The value of $n$ depends on the sample quality, temperature and sample size. There is a wide transition for $H_t < H < H_i$. For some of the samples, however, the experimental points are linearized in this field range, as well (inset to Fig.2). Finally, the second peak occurs at the fields around $H = H_i$, followed by a large data scatter for larger fields (experimental points with large data scattering for $H > H_i$ were deleted in Fig.2).

The values of $H_t$ and $H_i$ are dependent on the sample quality and temperature. What is more interesting, they show a strong dependence on the sample size. This points to a connection between these fields and the dependence of $J_{c0}$ on the sample size [4], where $J_{c0}$ was calculated from $\Delta m$ and the size of the sample. Fig. 3 shows the temperature dependence of $H_t$ for the HIP-ed MgB$_2$ sample, whose volume was decreased after the measurements were performed at all the temperatures. Apparently, $H_t$ decreased with the size of the sample for all temperatures. Fig. 3 also shows the temperature dependence of $H_t$ for commercial 325 mesh MgB$_2$ powder (particles smaller than 44 μm). The values of $H_t$ were much smaller than for the HIP-ed samples. The commercial powder was then ground, to obtain particles smaller than about 20 μm. The value of $H_t$ was substantially decreased by grinding and was not observed any more at 30K. The values of $H_t$ and $H_i$ were generally higher for samples of better quality. For example, the MgB$_2$ wires with magnetic $J_{c0}$ of 112 and 356 kA/cm$^2$ had $H_t$ at 20K of 0.4 and 0.8 T, respectively.

The change of the value of $H_t$ with the sample volume, $V$, is shown in Fig 4 for the HIP-ed MgB$_2$. These samples were of rectangular shape, with the field aligned along the largest dimension. Their exact dimensions are given in Reference [4]. $H_t$ is proportional to ln($V$) for all the temperatures: $H_t = A + B \ln(V)$. The values of $A$ were 6.90, 4.95 and 2.19 T and the values of $B$ were 0.35, 0.25 and 0.11 T, for $T$ = 10, 20, and 30 K, respectively. For calculating these values, the sample volume was in mm$^3$ and field in Tesla. The experimental data in Fig.4 can be extrapolated to $H_t = 0$, which is a characteristic volume below which the inflection in $\Delta m$ vs. $H$ at $H = H_t$ does not occur any more. The data for all three temperatures seem to extrapolate to the same volume at $H_t = 0$, which we denote by $V_0$. From this, it follows for $V > V_0$:

$$H_t = B \ln \frac{V}{V_o} , \qquad (1)$$

where $V_0 = e^{-B/A} \approx 3 \times 10^{-9}$ m$^3$. The sample thickness (its smallest dimension) extrapolates to $c_0 \approx 60$ μm at $V = V_0$.

The variation of $H_t$ with sample size in Fig. 4 was obtained with changing all three dimensions of the sample simultaneously. To test whether $H_t$ is affected by the change of only one dimension of the sample, another set of measurements was performed for two groups of cylindrical samples. In the group D, the diameter of the sample was changed and its length was kept at 3.91mm [4]. In the group Z, the length of the sample was changed, whereas the diameter was kept at 1.54 mm. The field was aligned along the cylindrical $z$-axis for both the groups. These samples were the $MgB_2$ cores extracted from an iron-sheathed wire, but they had $J_{c0}$ comparable to the HIP-ed samples. The HIP-ed samples were of much higher hardness than the samples from the wire and they were too hard to file down to the small diameter without breaking. Fig. 5 shows that $H_t$ changes logarithmically with the sample length. It extrapolates to $H_t = 0$ at the sample length $Z_0 \approx 20$ μm. Inset to Fig. 5 shows a logarithmic change of $H_t$ with the sample diameter, as well. $H_t$ extrapolates to zero at the sample diameter $D_0 \approx 70$ μm. Therefore, $H_t$ extrapolates to zero at a characteristic length that is of the same order of magnitude, regardless of which of the dimensions is changed (Figs. 4 and 5).

Fig. 6 shows the change of $H_i$ with the sample volume for the HIP-ed samples. $H_i$ also changes with $V$ logarithmically, and it extrapolates to zero at the volume of the order of $10^{-6}$ mm$^3$. This corresponds to the smallest dimension of the sample of the order of a micrometer. Therefore, the characteristic length associated with the sample size dependence of $H_i$ is one order of magnitude smaller than the characteristic length associated with $H_t$. It is worth mentioning that the irreversibility field is usually defined by a small value of $J_c$, of the order of 100 A/cm$^2$. This field is for $MgB_2$ at the step in the plot of magnetically determined field dependence of $J_c$ (Fig. 1) [8, 9], very close to $H_i$. Apparently, the irreversibility field was also reported to change with the sample size [4].

The characteristic lengths associated with the transitions in the field dependence of $\Delta m$ at $H_t$ and $H_i$ indicate on the existence of superconducting screenings contributing to $\Delta m$ that circulate on these two length-scales, in addition to the screening around the whole of the sample. This screening is expected to be seen in the magneto-optical image (MOI) of the sample. MOI for HIP-ed rectangular samples showed a typical roof-type field profile as the field increased, corresponding to the overall field penetration into the sample (Fig. 7). The zig-zag lines are magnetic domain walls of the indicator film that are formed above the largest field gradients of the sample. On the top of this was a finer detail, showing an arrangement of round structures reminiscent of voids, scattered across the sample. The diameter of the round structures was about 20 μm. These structures formed irregularly shaped cells between them, of the average size of about 35 μm. The resolution of the MOI set-up did not allow us to check if there is another tier of voids in these cells, on a ten times smaller length-scale.

The optical image of the sample is shown in Fig. 8. A pattern of dark voids is an apparent feature of this image. The voids surround the cells of $MgB_2$, exhibiting a golden shine. Comparing the MOI with the optical image of the sample, it is apparent that the round structures observed in MOI correspond to the voids in the sample, surrounding the volume of shiny high-quality $MgB_2$ cells (Fig.8). These cells are connected via narrow bridges. The size of the $MgB_2$ cells is of the same order as the characteristic screening lengths $c_0$, $Z_0$ and $D_0$. A similar structure of the voids was observed for all other samples, too, including the pellets and iron sheathed wires. However, the shape and density of the voids were sample dependent.

IV. DISCUSSION

The existence of superconducting screening on the length-scales smaller than the sample size, which results in the inflexions in the field dependence of $\Delta m$, is reminiscent of the weak links in high-temperature superconductors. For the high-temperature superconductors, the screening on the smaller length-scale occurs as the weak links cease conducting current at elevated fields and superconducting screening is confined to disconnected superconducting grains. However, the current transport through weak links is negligible in high-quality $MgB_2$ [1]. Superconducting grains of $MgB_2$ are well connected, giving high value of $J_c$ and much weaker decrease of $J_c$ with the field than for the high-temperature superconductors. Therefore, such decoupling of the grains cannot occur in $MgB_2$.

A. Model with two superconducting screening lengths

The clues for understanding the inflexions in the field dependence of $\Delta m$ can be found in Figs. 7 and 8. The only anomalies in MOI on the length-scale corresponding to the inflexion at $H_t$ (i. e. of the order of $c_0$) are the round structures, which can be identified as the voids in the optical image (Fig. 8). The voids surround the cells of $MgB_2$, connected by narrow bridges. Superconducting screening on the length-scale of the whole sample can occur only if the screening currents flow through these bridges between the bulkier cells. The screening currents are concentrated in the bridges and the current density in them is larger than in the cells. The maximum persistent current density in the bridges is $J_c$. Because of this, the current density inside the cells that belongs to the overall screening of the sample is smaller than $J_c$. This allows for additional screening on the length-scale of the cells between the voids, so that the net current density in the cells also becomes $J_c$ (Fig.9). The current density for the screening of the cells will be denoted as $J_{cc}$ and the average current density of the overall screening of the sample as $J_{cs}$. For simplicity, we assume that the cells consist of a homogeneous material and neglect the screening on an even smaller length-scale, associated with $H_i$ (Figs. 1, 2 and 6).

The values and field dependence of $J_{cc}$ and $J_{cs}$ depend on the structure and size of the voids and $J_c$ of the $MgB_2$ matrix. However, they are also affected by an interaction between $J_{cc}$ and $J_{cs}$. Namely, $J_{cc}$ is either added or subtracted from $J_{cs}$ in the cells, depending on the position in the cells (Fig.9). The spatial distribution and values of $J_{cc}$ and $J_{cs}$ are such that the total superconducting screening of the sample, including the overall screening and screening of the cells, is the most efficient. Obtaining the values and spatial distribution of $J_{cc}$ and $J_{cs}$ theoretically is a complex problem. However, the values and field dependence of $J_{cc}$ and $J_{cs}$ can be obtained from experiment, using a simple model of superconducting screening on two different length-scales.

The field dependence of the measured $\Delta m$ can be modeled in the same way as the dependence of the zero-field $\Delta m$ on the sample size [4, 15]. For a cylindrical sample with diameter $D$ and volume $V$ [15]:

$$\Delta m(H) = [afJ_{cc}(H) + DJ_{cs}(H)]2V/3, \qquad (2)$$

where $a$ and $f$ are the size of the cells and the filling factor for the cells, respectively. For the samples of different shape, the only difference is the proportionality factor outside the square bracket. The contribution of the screening on the length-scales smaller than the cells is neglected for simplicity. This contribution would have an even smaller pre-factor than $af$ in Eq. (2). Neglecting the third screening is justified experimentally by the small value of $\Delta m$ for $H > H_i$ (Fig. 2), where the screening on the smallest length-scale is expected to dominate in $\Delta m$.

The form of the field dependence of $J_{cc}$ and $J_{cs}$ can be determined from Fig.2. The experimental points are linearized in Fig. 2 for $H < H_t$. For higher fields, there is a non-linear transition. However, linear parts were sometimes clearly observed for $H_t < H < H_i$ (inset to Fig. 2), depending on the sample and its size. The fact that there is a transition between two different field dependencies of $\Delta m$ at $H_t$, which in turn depends on the sample size, implies that one of the contributions in Eq. (2) dominates the *field dependence of $\Delta m$* below $H_t$ and the other one above $H_t$. Therefore, the fit of experimental points in each of the field ranges, away from the transition at $H_t$, should reveal the functional dependence of each of the screenings in Eq. (2) on the field. The linear part in Fig. 2 gives: $-\log[-d(\ln(\Delta m))/dH] = \Omega + \Xi \log[H]$. The general form for the field dependence of $\Delta m$ in this field range is obtained from this expression as: $\Delta m = \text{const.} \ast \exp(-(H/H_i)^n)$. This function is commonly called the stretched exponential, or Kohlrausch, Williams and Watts' function [5, 6]. We choose a general expression for $\Delta m$ as:

$$\Delta m / V = \alpha \exp\left(-\left(\frac{H}{H_1}\right)^{n_1}\right) + \beta \exp\left(-\left(\frac{H}{H_2}\right)^{n_2}\right), \tag{3}$$

without assigning $J_{cs}$ to a particular contribution in Eq. (3). The pre-factors $\alpha$ and $\beta$ contain the size of the cells and sample in Eq. (2). We stress that the dominance in Fig. 2 of one of the screening currents signifies that they decrease with the field much faster than the other currents and not that they necessarily give a dominant contribution into $\Delta m$. However, the field dependence of $\Delta m$ for each of the screening currents is still proportional to the length-scale on which they flow, i.e. *a* and *D*.

The selection of Eq. (3) for fitting the field dependence of $\Delta m$ was first tested by choosing arbitrary values of the fitting parameters. Plotting $-d(\ln(\Delta m))/dH$ vs. $H$ with log-log scales, a peak similar to the one in Fig. 2 was obtained. Lowering the value of $\beta$ and keeping all other parameters in Eq. (3) constant, the peak shifted to lower fields, which is consistent with the observed size dependence of $H_t$ (Figs. 4 and 5).

The fitting of the field dependence of $\Delta m$ with Eq. (3) was performed by first obtaining the values of $n_1$ and $H_1$ from the measured data. The gradient of the linear part of experimental points in Fig. 2 for $H < H_t$ is equal to $n_1-1$. Approximate value for $H_1$ was obtained by neglecting the second part of Eq. (3). Then, $\log[-d(\ln(\Delta m))/dH] \approx \log[n_1/H_1^{n_1}] + (n_1-1)\log[H]$. For $H = 1$ T in Fig. 2, $-d(\ln(\Delta m))/dH \approx n_1/H_1^{n_1}$. The value of $-d(\ln(\Delta m))/dH$ at 1 T was obtained by extrapolating the linear part in Fig.2 to 1T, which gave us the value of $H_1$ using the known value of $n_1$. This resulted in a very good fit for the field dependence of $\Delta m$ for $H < H_t$, with $\alpha$ being the only arbitrary fitting parameter. Therefore, the first part of Eq. (3) was chosen to fit the data for $H < H_t$.

The values of $H_2$ and $n_2$ could only be obtained for several samples, where the experimental data were linear for fields $H_t < H < H_i$ (inset to Fig.2). This was only possible where a combination of the parameters in Eq. (3) resulted in sharp transitions at $H_t$ and $H_i$, so that the transitions did not distort the linearity of the experimental points between $H_t$ and $H_i$. For these samples, it was observed that the value of $H_2$ was close to 2.5 T and it was always larger than $H_1$. The value of $n_2$ was about 2.6. These were the starting values for fitting the field dependence of $\Delta m$ with Eq. (3) for the other samples. They were refined, so that the fit closely matched the experimental data. Adding the second part of the field dependence of $\Delta m$ into the fitting expression, Eq. (3), required a significant change in the value of $\alpha$. However, the values of $n_1$ and $H_1$ required only a small change.

The fit of the field dependence $\Delta m/V$ for the HIP-ed sample with $V = 0.23$ mm$^2$ is shown by the solid line in Fig. 10. Apparently, Eq. (3) describes the field dependence of $\Delta m$ very well. Equally good fitting was obtained for all the samples measured. The experimentally observed logarithmic dependence of $H_t$ on the sample volume (Eq. (1) and Figs. 4 and 5) was also obtained from the fits using Eq. (3). Dashed and dotted lines show separately the first and second part of Eq. (3), respectively. The value of the zero-field $\Delta m$ for the fit with the first part of Eq. (3) is about two times higher than for the second part. However, the former exhibits much stronger field dependence than the latter. Because of this, $\Delta m$ at elevated fields ($H \gg H_t$) is contributed only by the screening described by the second part of Eq. (3) (dotted line in Fig. 10). On the other hand, the second part of Eq. (3) is almost constant for $H < H_t$. Therefore, the field dependence of $\Delta m$ for $H < H_t$ is defined only by the screening described by the first part of Eq. (3). Nevertheless, both parts of Eq. (3) give a significant contribution to $\Delta m$ for $H < H_t$. The relative contribution of each of the parts into $\Delta m$ depends on the sample size when the sample size approaches the value of $c_0$, whereas it does not change any more with the sample size for much larger samples. This is in agreement with the reported artificial sample size dependence of magnetically determined $J_{c0}$ for small samples, if only the overall screening of the sample was assumed [4].

B. Stretched exponential field dependence of $J_c$

An interesting outcome of our measurements is that the field dependence of $J_c$ is described by the stretched exponential function, instead of the commonly used exponential function. The exponential function would give a constant value in Fig. 2, which is by far different from the experimental data. Further, $H_t$ is the field for which $d^2[\ln(\Delta m)]/dH^2 = 0$. This condition gives a complicated equation which for the exponential function (i.e. $n_1 = n_2 = 1$) becomes: $H/H_1 = H/H_2$. The only non-trivial solution for this equation is $H_1 = H_2$, which is simply a single exponential function that does not exhibit the transition observed experimentally at $H_t$ (Figs. 1 and 2). Further, simulations with the combination of two exponential functions with $H_1 \neq H_2$ could not produce the peak at $H_t$ in Fig. 2.

In our fitting of $\Delta m$ vs. $H$, the values of both $H_1$ and $H_2$ changed with the sample size. For the HIP-ed samples, the values of $H_1$ and $H_2$ at 20K were in the range of 1 - 1.9 T and 2.2 – 2.8 T, respectively. The values of $n_1$ and $n_2$ were in the range of 2.2 – 3.2 and 2.4 – 2.9, respectively.

It is not clear why exactly the field dependence of $\Delta m$ follows the stretched exponential function. This function is commonly used for describing the structural relaxation of glasses [16, 17, 18] and supercooled fluids [19, 20], relaxation of remanent magnetization of spin glasses [21, 22, 23], dielectric relaxation [6], and many others. In addition to its significance as a common phenomenological relaxation function, it was found theoretically that the relaxation of a fractal system with closed configuration space is necessarily of the stretched exponential type [24]. Our measurements, however, show that the *field dependence* of $\Delta m$ for $MgB_2$ is a stretched exponential. The same was found earlier for both, transport and magnetic $J_c$ of high-temperature superconductors [25]. Both, $MgB_2$ and high-temperature superconducting samples have porous structure, with percolative current flow through them. Further, the value of the parameters $n_i$ and $H_i$ in Eq. (3) depends on the sample size as it approaches the size of the basic screening cells in that sample. This would all suggest that the stretched exponential form of $\Delta m$ is associated with the percolative current flow in the sample.

C. Transport and magnetic $J_c$

It is important to distinguish the part of the Eq. (3) that corresponds to $J_{cs}$. This is because $J_{cs}$ is the average critical current that screens the whole of the sample, which is equivalent to the transport critical current. The field dependence of $\Delta m$, obtained from the transport measurements of $J_c$ and back-calculated using the critical state model, is shown in Fig.10 (solid symbols) for comparison with the magnetic measurements. The transport measurements were obtained for iron-sheathed wires, with the field parallel to the wire. It was shown that the effect of the iron sheath on the results is then minimized [26]. Considering that the measurements for two different samples are compared, there is a remarkable agreement between the $\Delta m$ obtained from transport measurements and the fit with the second part of Eq. (3). Therefore, the field dependence of $J_{cs}$ is defined by $\beta$, $H_2$ and $n_2$ in Eq. (3). Further, the first part of Eq. (3) cannot describe the transport measurements because it decreases very strongly with field for $H > H_t$, in contrast to the transport measurements that still give large values of $J_c$ for these fields. Consequently, the field dependence of $J_{cc}$ is described by $\alpha$, $H_1$ and $n_1$ in Eq. (3).

All the results of this paper show that the field dependence of $J_c$ cannot be obtained by simply applying the critical state model to $\Delta m$ obtained from magnetic measurements. Due to the inhomogeneous structure of $MgB_2$, $\Delta m$ is contributed to by superconducting screenings on different length-scales and each of them has different stretched exponential field dependence. Because of different values of $n_i$ and $H_i$ in Eq. (3), each of the screening gives a dominant contribution to the field dependence of $\Delta m$ in a particular band of the fields. At low fields, the largest contribution comes from the screening currents around the cells between the voids in all types of samples, including the pellets, powders, and iron-sheathed wires. In addition, the contribution of these screening currents becomes even more dominant for $H < H_t$ as the sample size becomes smaller than a few millimeters. Because these currents only flow on the length-scales smaller than 0.1 mm, $J_c$ calculated from $\Delta m$ and the length scale of the size of the sample will be artificially several times higher than the transport $J_c$ for $H < H_t$ (Fig.10).

For $H_t < H < H_i$, the screening of the whole of the sample gives dominant contribution to $\Delta m$ (Fig.10). The contribution from the cells, and additional third contribution that dominates for $H > H_i$ and was neglected in the above discussion, is insignificant in this field range, away from $H_t$ and $H_i$. The $\Delta m$ for $H_t < H < H_i$ can be used in the critical state model to calculate $J_c$, which is equivalent to the transport $J_c$.

The transition at $H_i$ indicates the existence of a third tier of superconducting screening. The sample size dependence of $H_i$ is qualitatively the same as for $H_t$ (Fig. 6), indicating the same underlying principle for both the transitions ( Eqs. (2) and (3)). However, Fig. 6 shows the length-scale of the screening for $H > H_i$ is of the order of a micrometer. MOI set-up could not provide such a high resolution, however SEM examination provided the necessary clues. Fig. 11 a) shows the structure of the voids of the HIP-ed sample, also observed with optical microscope (Fig.8). Further magnification (Fig. 11 b)) reveals a grain-like structure of the size of about a micrometer. These are the clusters of superconducting crystals, which had already been identified as the cause of the step in the field dependence of $\Delta m$ [8]. The size of these clusters depends on the preparation procedure. Their size was about 200 μm in our first samples [8], however they were of the order of micrometer in our better quality samples. TEM work on high-quality $MgB_2$/Fe wires showed that there were actually two tiers of sub-micrometer clusters, of the size of tens of nanometers and almost a micrometer[27].

Our results (Figs. 6 and 11) imply that the micrometer-size clusters are responsible for the screening at $H > H_i$ in the HIP-ed samples. This is in agreement with our earlier results on poorer quality bulk $MgB_2$ [8]. Unfortunately, the field dependence of $\Delta m$ for the screening on the micrometer-scale could not be obtained accurately, due to the small value of $\Delta m$ at these fields. In rare measurements with low noise, it seemed to be following the exponential law for some samples, yet for others (Fig.1) it was more like the stretched exponential. Because these measurements are on the resolution limit of the instrument, it is difficult to distinguish between them reliably. The TEM images in Ref. [27] would imply that there is another screening, on the ten-nanometer scale. However, the detection of this screening was not possible, because the screening on such a small length-scale would give $\Delta m$ that is even smaller than for the screening on the micrometer scale.

D. Other models implying the sample size dependence of $\Delta m$

Measurements of magnetic relaxation for bulk $MgB_2$ samples of different thickness[28] showed that the pinning potential increases with the sample thickness for thin samples and it becomes constant for samples thicker than about 1 mm. This was explained by a model based on vortex rigidity. Because increased pinning potential is associated with larger $H_{irr}$, this model might also explain our results on the sample size dependence of $H_i$ (Fig. 6), as the value of $H_{irr}$ is close to $H_i$. In the proposed model [28], the pinning potential increases with the sample thickness for samples thinner than the collective pinning length in the direction parallel to the field [29]: $L_c \approx (\varepsilon^2 \xi^2/\gamma)^{1/3}$. Here, $\varepsilon = [\Phi_0/(4\pi\lambda)]^2$ and $\gamma$, $\xi$, $\lambda$ and $\Phi_0$ are the disorder parameter, coherence length, London penetration depth and magnetic flux quantum, respectively. It was assumed that $L_c \approx 1$mm. For samples thicker than $L_c$, the pinning potential was proposed to increase no further with the sample thickness, because the vortex lines break up into segments of the size $L_c$. However, using [30] $\xi \approx 10^{-6}$ cm, $\lambda \approx 10^{-5}$ cm and $\gamma \approx 1$, the estimated values of $L_c$ are orders of magnitude smaller than 1 millimeter. Because $H_i$ in our measurements increases for sample size of up to a few millimeters (Fig.6), it is unlikely that this model can explain our results. Further, the pinning potential does not change with the lateral dimensions of the sample in their model [28], i. e dimensions perpendicular to the field. However, the field dependence of $\Delta m$ in our measurements changes with either lateral or longitudinal dimension of the sample (Fig. 5).

The value of $J_c$ is defined by an arbitrary value of the electrical field ($E$) produced between the voltage contacts in transport measurements, or by the sweep rate of magnetic field in the magnetic measurements. The latter is equivalent to the former, because for a cylindrical sample with field along its $z$-axis, for example, $E = -D/2 * dB/dt$. Therefore, $E$ is proportional to the sample dimension perpendicular to the field (i.e. $D$), which implies that $J_c$ depends on this dimension, as well [31]. Because of this, $\Delta m$ should also depend on $D$, which could explain our results in the inset to Fig.5. However, the field dependence of $\Delta m$ is then not supposed to depend on the length of the sample, which is in contrast to the observed dependence of $H_t$ on the sample length (Fig.5). Therefore, the change of $E$ with the sample size cannot be the mechanism for the observed field and sample size dependence of $\Delta m$. In addition, the observed transitions at $H_t$ and $H_i$ cannot be described by this mechanism. Further, the activation energy for magnetic vortices in $MgB_2$ is too high to allow observation of this effect. Namely, $E$ increases abruptly at $J = J_c$ and changing the definition of $J_c$ to another value of $E$ would alter the value of $J_c$ very little.

Contribution of both, bulk and Bean-Livingstone surface pinning into $\Delta m$ was also put forward to explain the size dependence of $\Delta m$ at low fields [31]. This concept could in principle explain the occurrence of the transition in the field dependence of $\Delta m$ at $H_t$. Namely, the surface pinning should be effective only for [32] $H < \kappa H_{c1}/\ln\kappa$, where $\kappa = \lambda/\xi$ and $H_{c1}$ is the lower critical field. At higher fields, the bulk pinning would dominate. However, using $\kappa \approx 10$ and $H_{c1} \approx 20 - 50$ mT at 20K [30, 33], this transition is expected to occur at about 0.2 T at 20K. Experimentally, however, the transition occurs at about 2T and it depends logarithmically on the sample size (Fig. 4). Additionally, a characteristic length-scale associated with $H_t$ is about 50 μm (Figs. 4 and 5). However surface pinning is associated with length-scales of the order of 10 nm (i.e. $\lambda$ and $\xi$). Therefore, if the field dependence of $\Delta m$ were dominated by the surface pinning for low fields, $H_t$ would extrapolate to zero at a three orders of magnitude lower length-scale. Most importantly, apparent irregularities in the structure of bulk $MgB_2$ will result in rough sample surface on the length-scale of $\lambda$ [27]. This provides numerous gates for the vortex entry[34], rendering the surface pinning ineffective for bulk $MgB_2$ samples.

The effect of geometrical barriers on our results was irrelevant, because the HIP-ed samples had the same geometrical proportions. Further, the field was always directed along the longest dimension of the samples, additionally minimizing effect of any geometrical barriers. The only exception were the short samples in Fig. 5, where the sample diameter was constant and its length was decreased for each subsequent measurement. However, even these measurements gave the logarithmic dependence of $H_t$ on the sample size (Figs. 4 and 5), the same as for the other samples.

In contrast to these models, the existence of the transitions at $H_t$ and $H_i$ and their dependence on the sample size are naturally explained by the current percolating between the cells of $MgB_2$ in the sample (Fig. 8). The field dependence of $\Delta m$ on the sample length may seem a bit unusual, but the percolation of the current can account for this, too. With the field along the cylindrical $z$-axis of the sample, the $z$-component of the local screening current will not be zero, as in a homogeneous medium. Because of the voids in the samples, the current screening the whole of the sample has to cross the bridges between the cells, which make it locally flow in the direction of the cylindrical $z$-axis. This $z$-component of the current will flow in the negative $z$-direction through other bridges, so that the net current in $z$-direction is zero. However, the local $z$-component of the current contributes to the overall screening of the sample, enabling additional links between the superconducting cells. As the sample length approaches the size of the cells, some of these links are destroyed, which changes the relative contribution of $J_{cs}$ and $J_{cc}$ to $\Delta m$, and thus affecting the values of $H_t$. The importance of the local $z$-component of the current is also supported by transport measurements on $MgB_2$ wires [26], where the field dependence of $J_c$ was the same for the field parallel and perpendicular to the overall transport current.

V. CONCLUSIONS

The magnetic field dependence of $\Delta m$ was shown to consist of a sum of at least two stretched exponential functions. The experiments showed that each of the stretched exponential contributions is associated with a superconducting screening on a particular length-scale. Each of these functions gives a dominant contribution to $d(\Delta m)/dH$ in a particular field range. This occurs because the stretched exponential function changes little in low fields, and fast at high fields, and because each of the screenings has different parameters $n_i$ and $H_i$ in the stretched exponential form (Eq. (3)). The fields at which the transitions from the dominance of one of the functions to the other occur ($H_t$ and $H_i$) are identified as inflection points in the field dependence of $\Delta m$ in log-log plots (Fig.1). For $H < H_t$, the screening on the length-scale of tens of micrometers determines the field dependence of $\Delta m$. For $H_t < H < H_i$, the field dependence of $\Delta m$ is dominated by the screening on the length-scale of the sample, i.e. a millimeter. These screening currents correspond to the transport critical currents. For still higher fields, the screening on the length-scale of the order of 1 μm determines the field dependence of $\Delta m$.

These tiers of screening on different length-scales occur because of the inhomogeneous microstructure of bulk $MgB_2$, consisting of an array of voids. The voids encircle cells of $MgB_2$ material, of the size of ~ 10 μm, connected by narrow bridges. The currents screening the whole of the sample percolate through the bridges, defining the field dependence of $\Delta m$ for $H_t < H < H_i$. Because the currents screening the whole of the sample are concentrated in the bridges, additional screening occurs around the cells of $MgB_2$ between the voids (Fig.9). This screening on ~ 10 µm scale defines the field dependence of $\Delta m$ for $H < H_t$. The cells themselves consist of ~ 1 µm large clusters of $MgB_2$ crystals. The superconducting screening around these clusters on ~ 1 µm scale (i.e. inside the cells) is associated with the field dependence of $\Delta m$ for $H > H_i$. The existence of the clusters inside the cells implies that the screening around the cells on ~10 µm scale ($H < H_t$) is also of a percolative nature. The percolation of the current is probably responsible for the stretched exponential field dependence of $\Delta m$, instead of the generally expected exponential dependence. The stretched exponential field dependence of $\Delta m$ was also obtained for high-temperature superconductors [25], where the current transport is also of a percolative nature.

Such a complex structure of superconducting screening is a cause of errors when calculating $J_c$ by simply applying the critical state model to the measured $\Delta m$. The field dependence of thus obtained $J_c$ will be actually a contribution of different types of screening, each of them defining the field dependence of $J_c$ in different field ranges. The value of $J_c$ will also be wrong: the screenings occur at different length scales, whereas the critical state model assumes that they circulate the whole of the sample. Further, $J_c$ will artificially depend on the sample size as it approaches the size of the screening islands in the sample. The best one can do is to separate the $\Delta m$ for $H_t < H < H_i$ and calculate $J_c$ using the size of the whole sample. The value of this $J_c$, and its field dependence, are in good agreement with the transport $J_c$.

Acknowledgments
This work was financially supported by the Australian Research Council.

Figure 1: Field dependence of $\Delta m$ for a bulk $MgB_2$ sample at 20K.

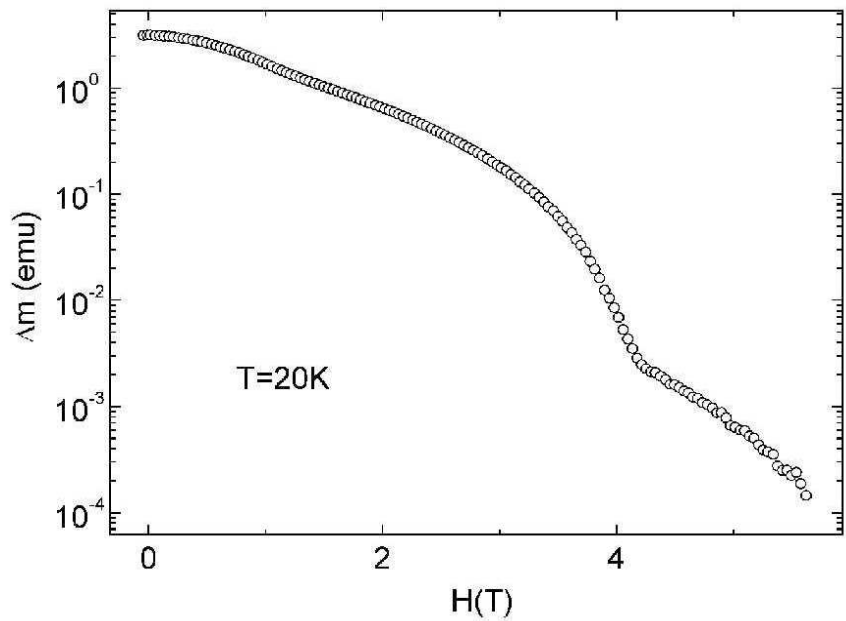

Figure 2: Field dependence of the relative decrease of $\Delta m$ with field, $1/\Delta m * d(\Delta m)/dH$, for the same sample as in Fig.1 at 20K.

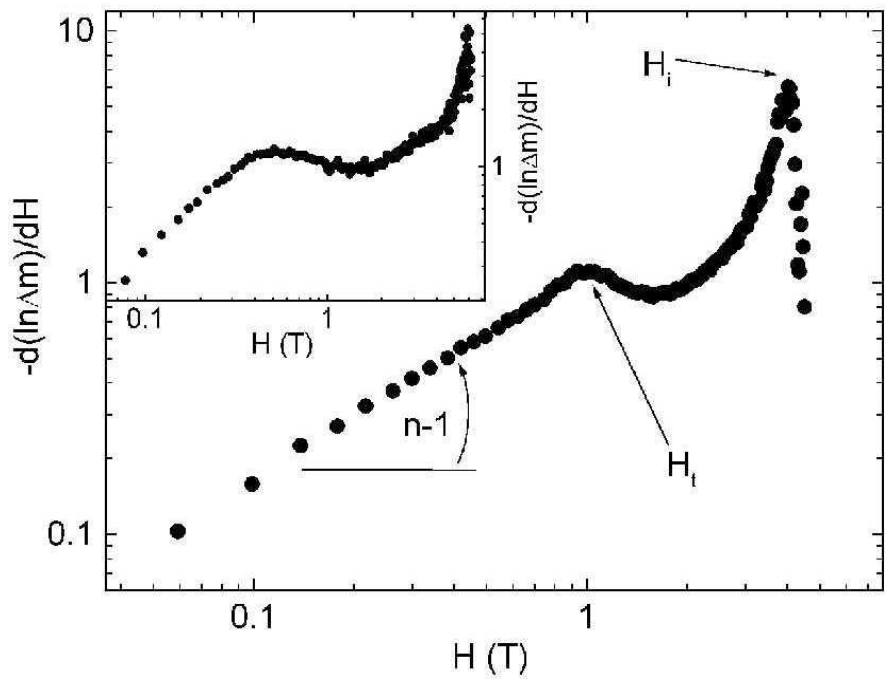

Figure 3: Temperature dependence of the transition field $H_t$ for several different samples: HIP-ed $MgB_2$ with volume of 12 mm$^3$ (open squares) and 0.23 mm$^3$ (solid squares), commercial $MgB_2$ powder with particle size less than 44 μm (solid circles), which was subsequently ground to obtain particles smaller than 20 μm (open circles).

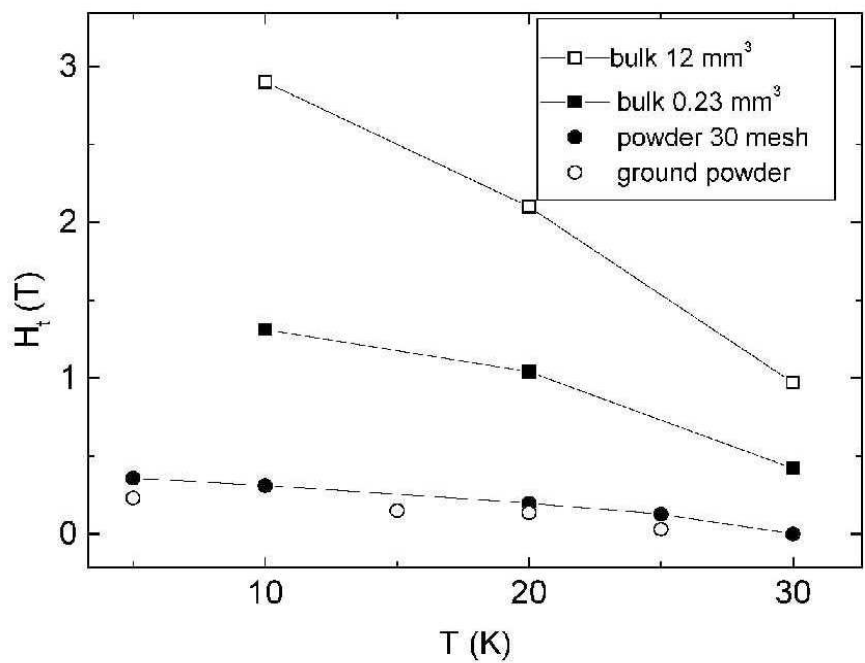

Figure 4: The dependence of the transition field $H_t$ on the volume of the HIP-ed $MgB_2$ sample, for $T$ = 10, 20 and 30 K. $H_t$ extrapolates to zero at $V = V_0$ for all three temperatures.

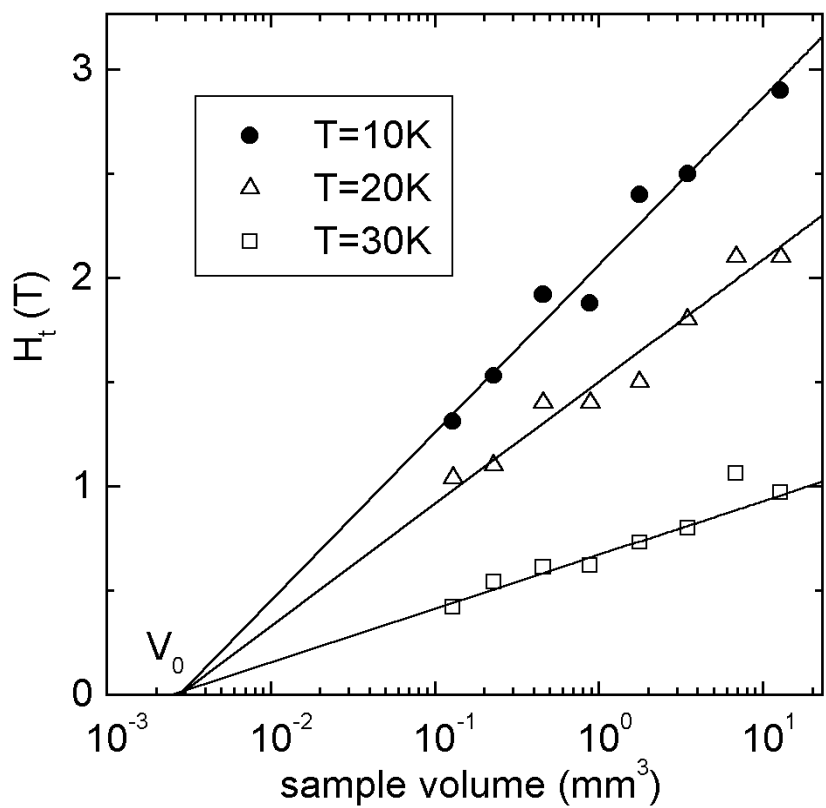

Figure 5: The dependence of the transition field $H_t$ on the sample length, for the core of an iron-sheathed $MgB_2$ wire. The diameter of the core was constant, 1.54 mm. Inset: The dependence of $H_t$ on the diameter of the core for an iron-sheathed wire, where the length of the core was kept constant, at 3.91 mm. The temperature was 20K.

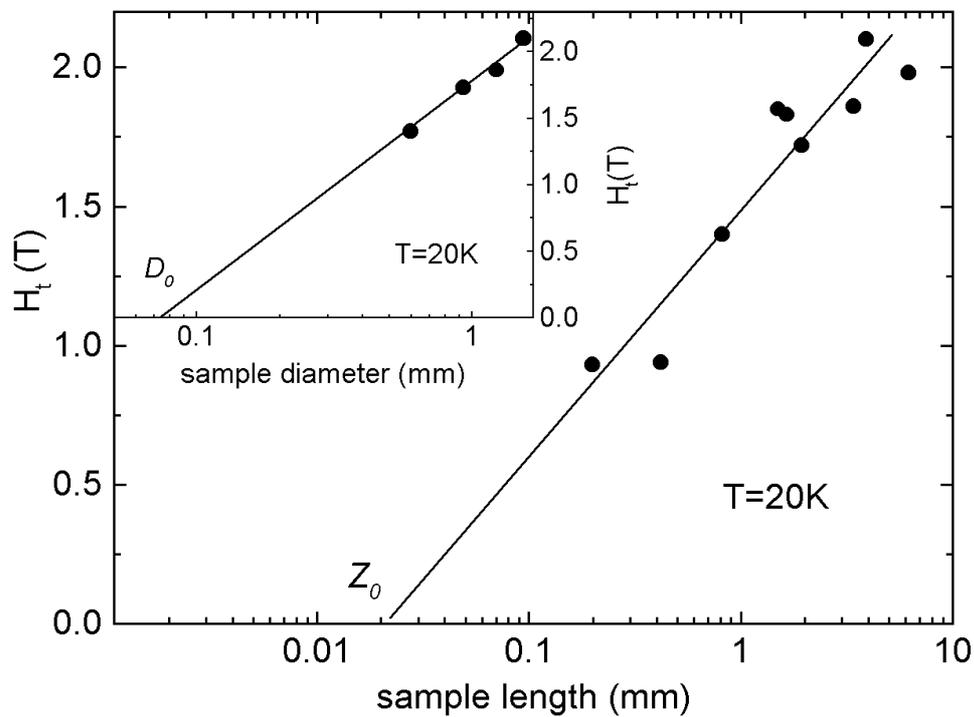

Figure 6: The dependence of the transition field $H_i$ on the volume of the HIP-ed $MgB_2$, for $T$ = 10, 20 and 30 K.

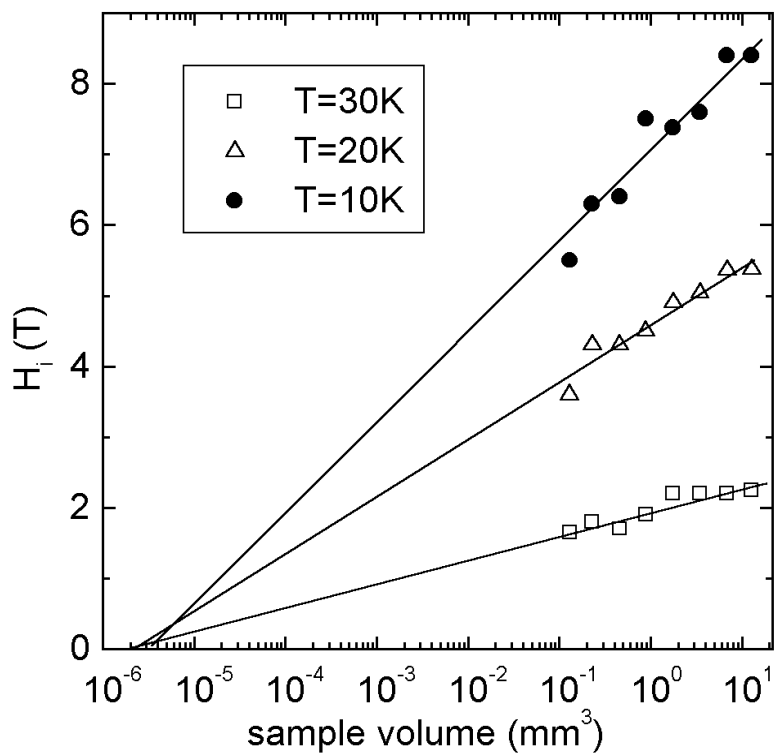

Figure 7: Magnetooptical image for a HIP-ed sample at 20 K. The bar represents 0.2 mm.

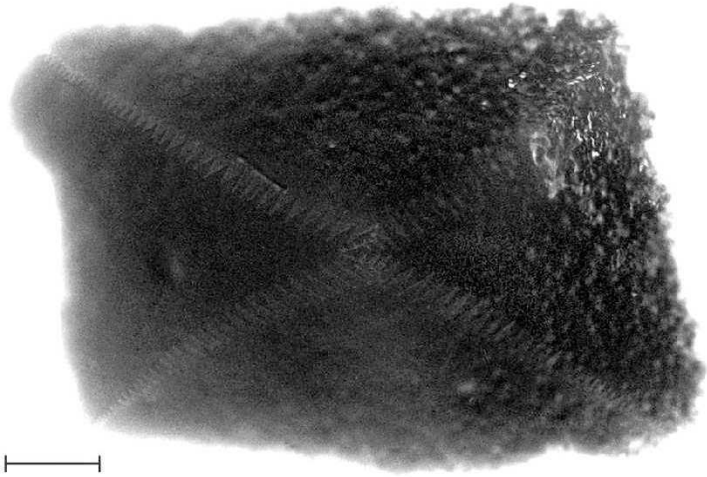

Figure 8: Optical image for the same HIP-ed sample as in Fig.7. The bar represents 0.2 mm.

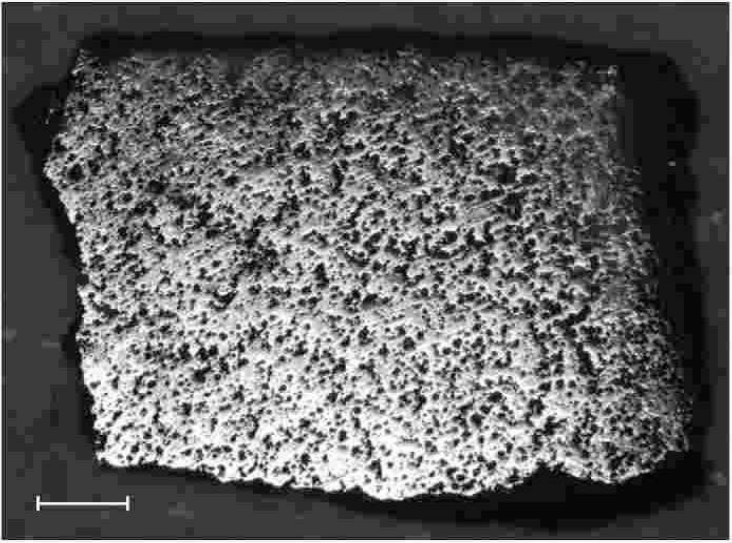

Figure 9: Schematic drawing of the screening currents in the sample. $J_{cs}$ and $J_{cc}$ are drawn by solid and dotted lines, respectively. The screening currents on 1 μm length-scale are represented by the dots. The shaded ellipses are the voids in the sample.

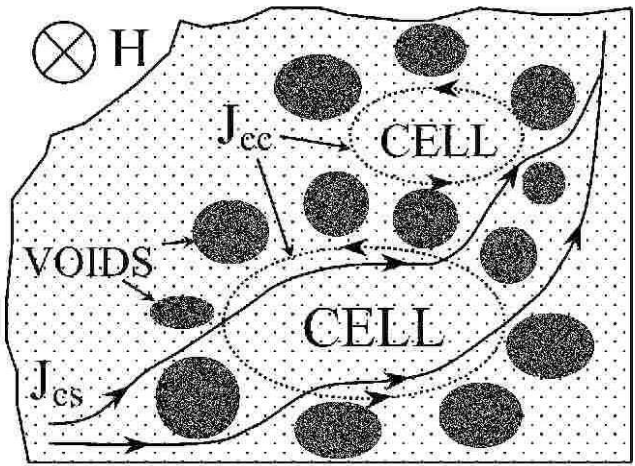

Figure 10: The field dependence of $\Delta m/V$ for the HIP-ed sample with volume of 0.23 mm$^3$, at 20 K. Solid line is the fit using Eq. (3), with: $\alpha = 912$ emu/cm$^3$, $H_1=1.17$ T, $n_1 = 2.32$, $\beta = 375$ emu/cm$^3$, $H_2 = 2.41$ T and $n_2 = 2.9$. Solid and dotted lines show separately the first and second part of the Eq. (3), respectively. Solid squares are $\Delta m/V$ calculated in the critical state model, with values of $J_c$ obtained from the transport measurements, for an iron sheathed MgB$_2$ wire at 20K.

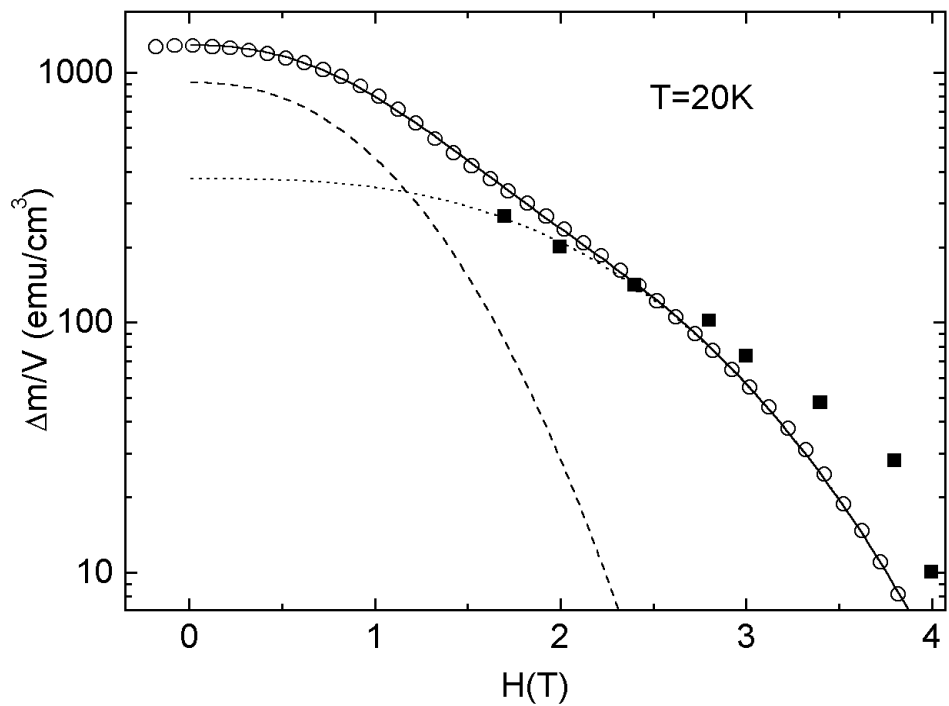

Figure 11: Scanning Electron Microscope image of a HIP-ed MgB$_2$ sample at two different magnifications. The sample was broken off a larger pellet, without polishing, to reveal the finer structure of clusters in the cells.

a)

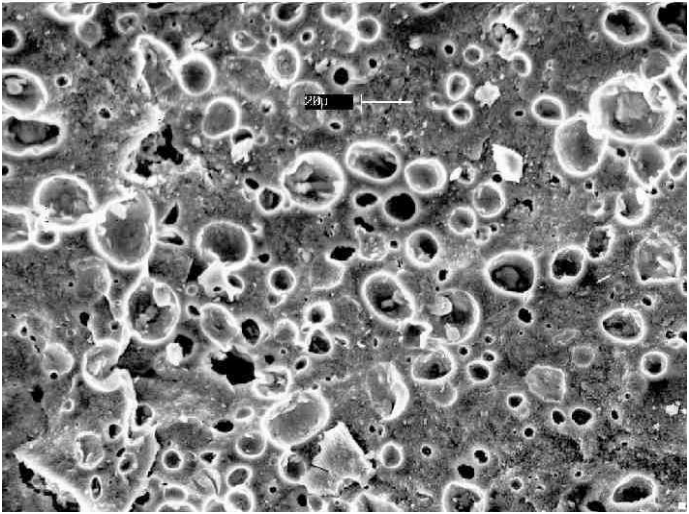

b)

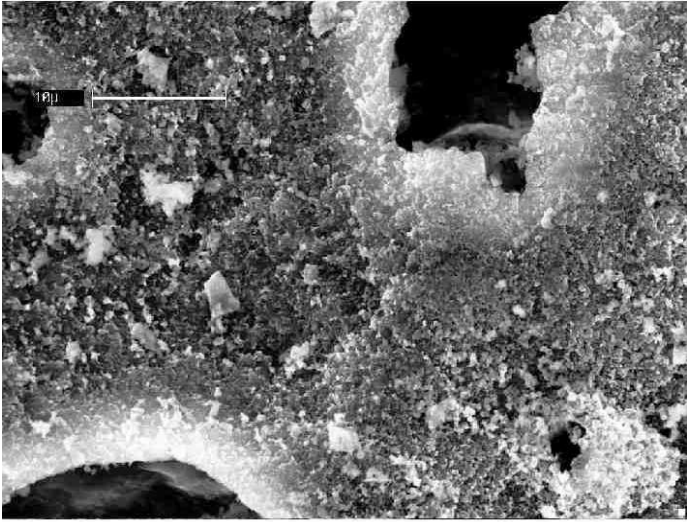